# XML Information Retrieval Systems: A SURVEY

Awny Sayed

Information Technology Dept. - Ibri College of Applied Science

Sultanate of Oman

Mobile Number 00968-98838296,

awny.ibr@cas.edu.om

**ABSTRACT:** The continuous growth in the XML information repositories has been matched by increasing efforts in development of XML retrieval systems, in large parts aiming at supporting content-oriented XML retrieval. These systems exploit the available structural information, as market up in XML documents, in order to return documents components- the so called XML elements-instead of the complement documents in repose to the user query. In this paper, we provide an overview of the different XML information retrieval systems and classify them according to their storage and query evaluation strategies.

**Keywords**: XML, XML storing, XML indexing, XML querying, Information Retrieval

## 1. INTRODUCTION

Indexing data for efficient search capabilities is a core problem in many domains of computer science. As applications centered on semantic data sources become more common, the need for more sophisticated indexing and querying capabilities arises. In particular, the need to search for specific information in the presence becomes of particular importance, as the information a user seeks may exist as an entailment of the explicit data by means of the terminology. This variant on traditional indexing and search problems forms the foundation of a range of possible technologies for semantic data. In unstructured information retrieval, it is usually clear what the right document unit is: files on your desktop, email messages, web pages on the web etc. While the first challenge in the semistructured information retrieving is that we don't have such a standard traditional document unit or indexing unit that is could be retrieved as a result to a query. The main profit of the XML which is considered as a new concept in the information retrieval branch is that when we query the XML documents we can dive deeply more than the document level allow to us into more specific units as document fragments (e.g. XML elements) which answer the user's query.

A new decision criterion that has been proposed for selecting the most appropriate and specific part of a document is the *structured document retrieval principle* [10]: *Structured document retrieval principle:* states that, {a system should always retrieve the most specific part of a document answering the query}. That principle motivates a retrieval strategy that returns the smallest unit that contains the information sought, but does not go below this level.

In our survey, we give an overview of the different XML information systems and classify them according to their storage and indexing strategies. For storage, we will answer the question, what is the best way of storing xml documents. Moreover, we will provide a classification of the different strategies used to store XML documents. The classification is based on the underlying system used for it (e.g., relational systems, object-relational systems, or native systems). For indexing and querying in our survey we will classify indexes into three parts (structured indexes, connection indexes, and path indexes) based on the underlying XML data, its tree-like structure or graph-like.

The rest of the paper is organized as follows. Section 2 introduces XML storage techniques. Sections 3 provide the details of the different indexing techniques; Finally, Section 4 concludes the paper and provides some suggestions for possible future research directions on the subject.

## 2. XML STORAGE TECHNIQUES

The basic properties of XML data are hierarchical tree-structured and semi-structured unlike ordinary relational databases. With this in mind in order to retrieve XML data efficiently we need different types of indexing techniques. An XML document can be modeled as a tree-like or a graph- like depending on the containment of that document to links or not. If the XML document does not contain such global or internal links it is modeled as a tree-like structure, otherwise if the XML document contains whether a global or internal links it is modeled as a graph-like structure. A tree, with nodes representing XML elements or attributes and edges representing parent-children relationships. Boxes with rounded corners represent attribute or text nodes.

### 2.1 Text Approach

The first strategy stores each XML document as a text file. One way to implement a query engine with this approach is to parse the XML file into a memory-resident tree against which the query is then executed. The tree is retained in memory as long as some nodes in the tree are needed for query evaluation. [23] found that the parsing time dominated query execution time and the approach was unacceptably slow. To make this approach competitive they adopted the following indexing strategy. Using the offset off an XML element inside the text file as its id, and build a path index mapping (*parent_offset, tag*) to *child_offset* as shown and an inverse path index mapping *child_offset* to *parent_offset*. These two indices are used to facilitate navigation through the XML graph. Another index mapping (*tagname, value*) or (*attribute_name, attribute_value*) to element offset is built to help evaluate selection predicates. A query engine can use these indices to retrieve segments of an XML file relevant to the query, reducing parsing time dramatically.

The main disadvantage of this approach is that whenever the XML document is updated, the element offset of preceding tags are also changed, which invalidates the indices and they have to be rebuilt. Regarding concurrency control it is necessary to lock both the XML document and the matching indices when some thread access data (reading/writing) due to data consistency. When a one thread is reading other threads can read as well, but when some thread is updating other threads cannot read or update the whole document since it cannot be considered consistent. The worst case is of course if new threads continue to access the document for reading, then it will not be possible to update any part of the document, unless some sort of prioritizing algorithm is implemented (and updates are given higher priority, of course this could lock out reads).

### 2.2 The Relational DTD Approach

The second strategy is the shared-inclining method proposed in and requires the existence of a Document Type Definitions (DTD). In DTD All element declarations begin with <! ELEMENT (case-sensitive) and end with >. They include the name of the element being declared followed by the content specification. In this declaration, the content specification is the keyword ANY (again case-sensitive). The element declaration <! ELEMENT SPEECH (SPEAKER, LINE+)> says that a SPEECH element must contain a single SPEAKER element followed by one or more LINE elements, the + quantifier indicates that the LINE must exist at least one time and no limits for the maximum number of its recurrence. An element that can only contain plain text is declared using the keyword #PCDATA in parentheses, like this: <! ELEMENT STAGEDIR (#PCDATA)> This declaration says that a STAGEDIR can contain only parsed character data, that is, text that's not markup. Like elements, the attributes used in a document must be declared in the DTD for the document to be valid. Attributes are declared by an attribute list in the following form: <! ATTLIST Element_name Attribute_name Type Default_value>.

A separate table is used to capture the set-containment relationship between an element and a set of children elements with the same tag. Each tuple in a table is assigned an ID and contains a parentlD column to identify its parent, an element that can appear only once in its parent is inline. If the DTD graph contains a cycle, a separate table must be used to break the cycle, the relational schema generated from the DTD and how the document is stored are shown below.

When reconstructing the XML document from this approach it is necessary to know how to build the document in terms of layout. Whether it is a partial or a full reconstruction does not matter because the work is the same, only when it is partial it is necessary to make specifications about which part one wishes to reconstruct. There is though a problem of recreating whitespace outside contents because this information is lost when the XML document is uploaded to the database.

## 2.3 Edge Approach

The third strategy is the "EDGE" approach described in  The directed graph of an XML file is stored in a single Edge table. Each node in the directed graph is assigned an *id* . Each tuple in the Edge table corresponds to one edge in the directed graph and contains the ids of the two nodes connected by the edge, the tag of the target node, and an ordinal number that is used to encode the order of children nodes. When an element has only one text child, the text is inlined.

TargetlD indicates that the edge points to a TEXT node or ATTRIBUTE node. 0 in ordinal field indicates an attribute edge. As suggested in an index is built on (tag, data)  in order to reduce the execution time of selection queries. We found that it was also very important to build indices on (sourceid, ordinal) and (targetID). The former is used to lookup children elements of a given element and the later is used when traversing from a child node to its parent.  The clustering strategy on the Edge table has significant impacts on query performance. While we clustered the Edge table on the Tag field, an alternative strategy is to cluster the table according to SourceID. This strategy has the benefit that sub-elements of one XML element are stored close to each other. The drawback of that Approach is that elements with the same tag name are not clustered. Consequently, queries such as "select all students whose major is Computer Science" will incur a large number of random I/Os. Similar to the EDGE model, the BINARY approach materializes the generic tree structure of XML documents in database tables. Hence, it is a model mapping approach as well

## 2.4 The Object Approach

An obvious way of storing XML documents in an object manager is to store each XML element as a separate object. However, since XML elements are usually quite small, all the elements of an XML document are stored in a single object with the XML elements becoming light-weight objects inside the object. [23] [24] use the term LW_object to refer to the light-weight object and file_object to denote the object corresponding to the entire XML document.

The offset of the lw_object inside a file_object is used as its identifier (lw_oid). The length field records the total length of the lw_object. The flag field contains bits that indicate whether this lw_object has opt_child, opt_attr, or opt_text fields. The tag field is the tag name of the XML element. The parent field records the lw_oid of the parent node. Opt_child records the lw_oids of the first and last child, if the lw_object has children. The sibling list of a node is implemented as doubly linked list via the prev and next fields. Opt attr  records the (name, value) pair of each attribute of the XML element. Text data is in-lined in the opt_text field if the text is the only child of the XML element; otherwise, the text data is treated as a separate lw_object. [23] built a B-Tree index that maps (tag, opt_text) and (attr_name, attr_value) to lw_oid. An element is entered in this index even if the opt_text field is empty so that this index can be used to retrieve all XML elements with a specific tag name. They also built a path index those maps (parent_id, tag) to child lw_oid. This optimized object approach is hard to perform concurrent operations on since the locking has to occur on the object representing the whole document; unless there should be build some extra concurrency control into the lw_objects themselves, but this would be overkill. To when locking anything in this approach means at least locking the whole XML document.

## 2.5 Native XML Storage Approach

Finally, we should have a look shortly at so-called native XML databases, which are specialized to store and process XML documents. Native storage schemas aim at efficient support for loading and storage complete documents as well as efficient navigation in documents. A native XML storage system store XML documents as flat files, i.e., it uses a

t*ext-based mapping*. However, evaluation of queries requires reconstructing the complete XML documents, which is not efficient when only parts of the documents are evaluated by the given query. As a result, most native XML storage schemas store XML documents as a tree structures based on the tree data model of XML [12] . These particular approaches are model-mapping approaches. Usually, native XML storage systems rely on the DOM tree representation of XML documents.

## 3. INDEXING TECHNIQUES

Since the hierarchical nature of the XML documents there is a lot of interesting in a query processing on data that conforms to a labeled- tree or labeled- graph model. To summarize, the structure of such data in the absence of a schema and to support path expressions evaluation, several structure indexes have been proposed for semi-structure data described as follows

## 3.1 Structure Indexes

The structure index I (G) of a data graph G is a summary graph that preserves all the paths in the data graph but contains a fewer number of nodes and edges To summarize the structure of such data in the absence of a schema and to support path expression evaluation, novel structural indexes [14], [19] have been proposed for semi-structured data. Unlike a schema, structure indexes are not prescriptive and thus may change with any update. Generalizations of these structures have gained increasing attention recently, as flexible index structures for XML [9], [16], [18], and size and performance issues in the original proposals have been addressed [18]. Pre/post schema encoding XML tree-structure.
In addition, the ideas behind these structure indexes have been used as statistical synopses for estimating path expression selectivity [2],[20]. Moreover, the structure index proposed in[and [13] presents a database index structure designed to support path expressions evaluation on trees. It has the capability to support all XPath axes and start traversal from any arbitrary nodes in an XML document. Building the index takes O (|E|), and space consumption is O (|V|), where V denotes the number of nodes in the XML tree and E the number of edges. The main idea of this index depends on the numbering schema. It computes two numbers for each element name in the XML data tree, one representing the *pre-order* and the other representing the *post-order.* These numbers are the result of a depth-first search on the XML data tree. Starting with the root element, the *pre-order* numbers are assigned in the order in which the nodes are visited during this search. *The post-order* defines the order in which the nodes are left. The authors explain that XPath axes (like ancestor and descendant axes) can be evaluated using these numbers. This index based on the following property for evaluating path expressions: For any two given nodes A and B in the tree, an arbitrary node B is a descendant of a node A, if and only if this condition is satisfied:

$$pre(A) < pre(B) \text{ and } post(A) > post(B)$$

If we want to evaluate all descendants of a given node using this schema, then the result is the set of all nodes that satisfies the above condition.
The *pre-/post-order* approach can be determined in a constant time by examining the pre-and post-order variable of the corresponding tree nodes. The [22] stated that the drawback of this approach is its lack of flexibility in case of changes to the structure of the XML-document. That is, the pre-/post-order variables need to be recomputed for the number of tree nodes if any update into the tree whether a new node is inserted or an existing one is deleted.

## 3.2 Connection Indexes

A *connection index* is the index which supports the XPath axes that are used as wildcards in path expressions (*ancestors-or-self, descendants-or-self, ancestors,* and *descendants*). Labeling schemes for rooted trees that support ancestor queries have recently been developed in the following researches.

In [4] and [16] they present a tree labeling scheme based on two level partition of the tree, computed by a recursive algorithm called prune&contract algorithm. All these approaches are, so far, limited to trees. We are not aware of any index structure that supports the efficient evaluation of ancestor and descendant queries on arbitrary graphs. The one, but somewhat naive, exception is to pre-compute and store the transitive closure Cx = (Vx, $E_x^+$) of the complete XML graph Gx = (Vx ,Ex) $C_x$ is a very time-efficient connection index, but is wasteful in terms of space. Therefore, its effectiveness with regard to memory usage tends to be poor (for large data that does not entirely fit into memory) which in turn may result in excessive disk I/O and poor response times. To compute the transitive closure, time $O(|V|^3)$ is needed using the Floyd- Warshall algorithm. This can be lowered to $O(|V|^2 + |V|\cdot|E|)$ using Johnson's algorithm. Computing transitive closures for very large, disk-resident relations should, however, use diskblock- aware external storage algorithms. [1] [7] [8] implemented the "semi-naive" method [BR86] that needs time O ($|E_x'|\cdot|V|$). Although there are several approaches are proposed to evaluate all the ancestors of a given node and test the reachability between two given nodes. For example, labeling schema proposed in [17] is called a *prefix-labeling schema* to handle a dynamic XML tree. The nodes in the XML tree are labeled such that the ancestor relationship is determined by whether one label is a prefix of the other. New nodes can be inserted without affecting the labels of the existing nodes. They define an assignment of binary strings to the edges of the tree, such that, the collection of strings associated with the outgoing edges from any node is prefix free, *a prefix free assignment*. At the first, the simple prefix schema finds a prefix free assignment to the tree. Then, it is label every node *v* with the concatenation strings assigned to the edges of the path from the root node to *v*.

For every assignment, labels are unique. Node *u* is ancestor of node *v*, iff the label of *u* is a prefix of the label of *v*. One major problem related to this approach is how to find an assignment that minimizes the sum of the lengths of the labels, unfortunately this problem is NP-hard [17] means no optimal solution to this problem. The main goal of the work in [17] is to find an assignment that minimizes the maximum length of the labels by using Huffman's algorithm [14]. Several labeling schemes are proposed using the above technique, for example, [4] [6] proposed a labeling schema for rooted trees that supports ancestor queries by assigning to each node in the tree a label which is a binary string. Given the labels of two nodes *u* and *v* it can be determined in a constant time whether *u* is an ancestor of *v* only by looking at the labels. Another labeling schema proposed on [25], it takes the advantages of the unique property of prime numbers to meet this need. Answering the ancestor-descendant queries for a given two nodes by only looking at the labels (based on prime numbers). An analytical study of the size requirements of the prime numbers indicates that this schema is compact and hardly affected.

Moreover, the authors introduced several optimization techniques to reduce the size of the schema. Unfortunately, these indexing techniques were supposed to handle tree-structure data. Extension of these techniques to the context of graph data could be very difficult because of the possibly exponential number of paths in the graph. Moreover, it may require a lot of computing power for the creation process and a lot of space to store the index.

## 3.3 Path Indexes

A path index is the index which supports the navigational XPath axes (*parent, child, descendants-or-self, ancestors-or-self, descendants,* and *ancestors*). Recent work on path indexing is based on structural summaries of XML graphs. Some approaches represent all paths starting from document roots, e.g., Data Guide [14] and Index Fabric [11]. T–indexes [19] support a pre– defined subset of paths starting at the root. APEX [9] is constructed by utilizing data mining algorithms to summarize paths that appear frequently in the query workload. The Index Definition Scheme [16] is based on bisimilarity of nodes. Depending on the application, the index definition scheme can be used to define special indexes (e.g. 1–Index, A(k)–Index, D(k)–Index [QLO03], F&B–Index) where k is the maximum length of the supported paths. Most of these approaches can handle arbitrary graphs or can be easily extended to this end. Most of these indexes are quite efficient in evaluating simple path queries. These indexes widely differ in space utilization, support for paths with wildcards (wildcard means the arbitrary long paths from source point to targets in XML graph). These path indexes depend on the structure summaries of the XML graph. Structure summary is an important technique for indexing XML arbitrary graph, in case the general schema of the information is missing. Using this summary of the data, one can evaluate the path expression queries without looking at the original data. In the following, we will describe these indexes in details.

### 3.3.1 Data Guide

DataGuide [14] is a "structural summary" for semistructured data and may be considered as analog of traditional database schema in context of semistructured data management. The DataGuide is a descriptive schema for XML data. While prescriptive schemas (DTD, XML Schema, Relax-NG) act more as a traditional database schema, restricting allowable XML data, a DataGuide infers rather than imposes structure. DataGuide describes actual (rather than possible) structure of XML data extracting the structure from the XML data. It may be used as schema for semistructered data without any explicit schema declaration, such as non-validated XML documents.

The dataguide is based on the *Object Exchange Model* (*OEM*) which is the simple and flexible data model that originates from the *Tsimmis* project at Stanford University [PGW95]. OEM itself is not particularly original, and the work presented using OEM adapts easily to any graph-structured data model. A value may be atomic or complex. Atomic values may be integers, reals, strings, images, programs, or any other data considered indivisible. A complex OEM value is a collection of 0 or more OEM subobjects, each linked to the parent via a descriptive textual label. Note that a single OEM object may have multiple parent objects and that cycles are allowed. For more details on OEM and its motivation.

[14] Describes the DataGuide that it is, intended to be a concise, accurate, and convenient summary of the structure of a database. They assume that the source database is identified by its root object. To achieve conciseness, they specify that a DataGuide describes every unique label path of a source exactly once, regardless of the number of times it appears in that source.

To ensure accuracy, they specify that the DataGuide encodes no label path that does not appear in the source. In addition they require that a DataGuide itself be an OEM object so we can store and access it using the same techniques available for processing OEM databases.

### 3.3.2 Indexing Template-compliant Paths: T-index

Like *DataGuide* [14], *1-index* [19] is intended to be used by queries that search the database from the root for nodes matching some arbitrary path expressions. *1-index* therefore, represents the same set of paths from the root like *DataGuide*. The main idea behind the index construction is the generation of a *non-deterministic automaton (NFA)* [22] to get more compact structure than the *DataGuide*. To construct the *1-index* of a data graph, the authors compute for each node the equivalence class using a bisimulation as equivalence relation which is defined in the next definition.

***Definition 3-2*** *(Equivalence Relation "□"):* For each node $u$ in the data graph, let the set L$u$= {$w$ □ a path from the root to node $u$ labeled $w$}. The set L$u$ may be infinite when the graph has cycles; however, it is always a regular set. Given two nodes $u$ and $v$ in the data graph we say that they are *language-equivalent* in notation $u$ □$v$, if L$u$= L$v$.

***Definition 3-3*** *(Bisimilarity):* Two nodes in the data graph are bisimilar (□) if all label paths into them are the same. In other words, if node $u'$ is parent of node $u$, node $v'$ is the parent of node $v$. If the two nodes $u$ and $v$ have the same label, then, u □$v$ if u'□$v'$.

Using bisimulation to deal with the index size and the construction cost problems that *DataGuide*

index yields. Where the size of the *DataGuide* may be large as the database itself, while *1-index* is at most linear.

The advantage of *1-index* and its family (*2-index* and *T-index* [19]) is that, it can be used to evaluate any path expressions accurately without accessing the data graph. However, the size of *1-index* can be quit larger for irregular XML data. Moreover, not all structures are interesting and most queries probably only involve short path expressions.

A(k)-index*:* A(k)-index [18] is a type of approximate structural summary of data graph since it does not reflect whole structure and nodes of XML tree are grouped according to the local structure. With these properties in mind we can think of several issues as follows.
- Not all structures are interesting.
- Paths longer than k may be never used.
- Complex paths may never show up.
- Longer path results in large index graph, which takes time to construct and traverse while querying.

We can reach to one solution considering above issues, that is, use of local similarity, which is approximate structural summary. We focus on features of A(k)-index in the following sections within the view of implementation issues.

Taking advantages of local similarity [3], the *A(k)-index* can be substantially smaller than *1-index* [19]. The parameter *k* control the "resolution" of the entire A (k)-index; all index nodes have the same local similarity of *k*. If *k* is too smaller, the index cannot support long path expressions accurately. If *k* is too large, the index may become so large. At this case, evaluating any path expression over this index will be expensive. The time required to build the index is $O(km)$ where *m* is the number of edges in the data graph. Furthermore, not all path expressions of length *k* are equally common. The *A(k)-index* lacks the ability to make certain parts have higher resolution than the others do, so it can not be optimized for complex path expressions with wildcards.

D(k)-Index*:* The D(k)-index is an adaptive summary structure for the general graph-structured data proposed recently. It allows different index nodes to have different local similarity requirements that can tailored to support a given set of frequently used path expressions and to avoid the A(k)-index drawbacks. For parts of the data graph targeted only by longer path expressions, a larger k can be used for finer partitioning. For parts targeted only for shorter path expressions, a smaller k can be used for coarser partitions. However, as a generalization of 1-index and A(k)-index, the D(k)-index processes the adaptive ability to adjust its structure according to the current query loads. D(k)-index has a very nice property compared with 1-index and A(k)-index because of dynamics. The author provides an efficient algorithms to update the D(k)-index with changes in the source data . The general approach of the D(k)-index is flexible and powerful, but the index design still has several limitations that need to overcome. For example of these limitations, the construction procedure of the D(k)-index forces all index nodes with the same label to have the same local similarity, which is unnecessary and restrictive. The D(k)-index also proposes a promoting procedure that incrementally refines the index to support a given set of frequently used path expressions. This procedure increases the local similarity of an index node if it reached by a given set of frequently used path expressions in the index graph. This index node will be partitioned into smaller nodes, all with the same increased local similarity. However, the problem is that in general the index node to be refined also points to data nodes that are irrelevant to the given set of frequently used path expressions.

*Definition (Index Graph)***:** Index graph means that we reduced the graph that summarized all the paths from the root in the data graph, the nodes that have the same label from root are collected into one node called index node. The index graph is smaller than the data. Path expressions can be directly evaluated from the index graph and can retrieval label-matching nodes without referring to the original data graph.

M(k) Index**:** To overcome these limitations for the D(k)-index, A M(k)-index (for "Mixed-k") is proposed in [15]. The authors built on the strength of D(k)-index and proposed M(k)-index and M*(k)-index to overcome its limitation. To overcome the limitations of over-refinement of irrelevant index nodes and data nodes, M(k)-index is proposed to target only the data nodes relevant to frequent queries. Like the D(k)-index, the M(k)-index uses the k-bisimilarity equivalence relation but allows different k values for different nodes; it is also incrementally refined to support new frequently used path expressions extracted from the query workload. Unlike the D(k)-index, however, M(k)-index is never over-refined for irrelevant index or data nodes. Thus, the M(k)-index has a smaller size without sacrificing support for any frequent used path expressions. To overcome the limitations of over-refinement due to overqualified parents and single resolution each node, M*(k)-index is introduced as a collection of M(k)-indexes whose nodes are organized in a partition hierarchy, allowing successively coarser partitioning information to co-exist with the finest partitioning information required. The M*(k)-index maintains k-bisimilarity information for all k up to some desired maximum, which can be different across nodes and adjusted dynamically according to the query workload. This feature allows the M*(k)-index to avoid over-refinement due to overqualified parents and support both short and long path expression queries over the same data nodes at the same time. Experiments show that although keeping partitioning information at different resolutions requires extra storage space; it is negligible compared to the savings achieved by avoiding over-refinement. The performance gain from query processing further justifies this new approach.

## 4. CONCLUDING REMARKING

After reviewing a number of existing XML information and retrieval systems, we can draw some conclusions about the state of the art and general trends in the fields. Our Survey addresses what exactly are the requirements for *efficient XML storage management.* A storage management schema must cover the following aspects efficiently: lossless storage of XML documents, complete and efficient reconstruction of decomposed XML documents, and support for processing path expressions on the XML document structure, support for processing of precise and vague predicates on XML content, navigation in XML documents, and online updates of XML documents. Moreover, IR community applies with some modification standard IR techniques for focused element-level retrieval. But despite some similarities with unstructured text, XML needs special treatment in terms of relevance of its elements to a user query and in its evaluation. Hence we need a new paradigm in its retrieval techniques and evaluation metrics. On the whole, XML as an research area holds immense prospect which is not still extensively explored and therefore remains an interesting field of further research.

On the other hand, for Indexing and querying XML data our survey introduces a short classification of structures indexes for semistructured data based on the navigational axes they support. *Structure index* supports all navigational for XPath axes. *Connection index* supports the XPath axes that are used as wildcards in path expressions (ancestor (descendant)-or-self-relationship and ancestor-descendant relationship). *Path index* supports only the following kinds of XPath axes (parent-child relationship, ancestor-descendant relationship, ancestor-or-self relationship, and descendant-or-self relationship*)*.

For heterogeneous XML documents in the Web (divided XML documents into several subcollections), a single index structure may not be appropriate. Therefore, it will be investigated whether it makes sense to combine several indexes as building blocks. This would allow for building an index for each subcollection and evaluating the proposed queries by "navigating" through the underlying sub-collection only.

Moreover, The most common web technology that will realize Web 3.0 is RDF (resource document Framework) model. The Resource Description Framework (RDF) is a flexible model for representing information about resources in the web. With the increasing amount of RDF data which is becoming available, efficient and scalable management of RDF data has become a fundamental challenge to achieve the Semantic Web vision. So, the most important question is, could we apply the same technologies used to store and retrieval Xml to RRD Data, this is still a very hot topics for research.